\documentclass[doublecol]{epl2}
\usepackage{amssymb}
\usepackage{graphicx}
\usepackage{dcolumn}
\usepackage{amsmath}
\usepackage{bm}
\usepackage{epsfig}
\usepackage{eucal}

\title{Transition from small to large world in growing networks}
\shorttitle{Transition from small to large world in growing networks} 

\author{S. N. Dorogovtsev\inst{1,2} \and P. L. Krapivsky\inst{3} \and J. F. F. Mendes\inst{1}}
\shortauthor{S. N. Dorogovtsev \etal}

\institute{                    
  \inst{1} Departamento de F{\'\i}sica da Universidade de Aveiro, 3810-193 Aveiro,
Portugal\\
  \inst{2} A. F. Ioffe Physico-Technical Institute, 194021 St. Petersburg, Russia\\
  \inst{3} Department of Physics and Center for Molecular Cybernetics, Boston University, Boston, MA 02215, USA
}

\abstract{
We examine the global organization of 
growing networks in which a new vertex is attached to already existing ones with a probability depending on their age. 
We find that the network is infinite- 
or finite-dimensional depending on whether the attachment probability decays slower or faster than $(\text{age})^{-1}$. 
The network becomes  one-dimensional when the attachment probability decays faster than $(\text{age})^{-2}$. 
We describe structural characteristics of these phases and transitions between them. 
}

\begin{document}

\maketitle






A relation between small worlds (compact, infinite-dimensional objects) and
large worlds (finite-dimensional objects) is a key issue in understanding of
networks. The surge of interest in networks has been triggered by the paper of
Watts and Strogatz \cite{ws98} who found 
a transition from lattices to networks with a small-world organization. The Watts-Strogatz construction describes a smooth crossover between the two limiting situations. A sharp transition between 
finite- and infinite-dimensional networks 
(it coincides with a so-called searchability point) has been observed in studies of decentralized search algorithms
\cite{k00,k06,m02}. These findings have been made in ``static'' networks.  Results
of numerical simulations \cite{h07,hb06,tzs06,zzw07} suggest a sharp transition also in specific growing network models. The existence of this special point was confirmed 
analytically in work \cite{l07}. Yet essentially nothing is known about the nature of 
these transitions in static and especially in growing random networks. Here we provide an analytical description of a transition from 
an infinite-dimensional network to a finite-dimensional one 
in random recursive graphs with aging. 

A random recursive graph is one of the two principal models of a random graph (the second one is the Erd\H{o}s-R\'enyi model). A random recursive tree (a tree is a connected graph without loops) is built according to the following procedure --- a new vertex attaches equi-probably to an already existing vertex.  Random recursive trees are well understood, see e.g. Refs.~\cite{mh,d96,drmota,GNR,m74,s90,m91}. 
Here we generalize this model by letting the choice of the target vertex to be proportional to a certain function, $w(a)$, of the age. 
(The vertices in the network are labelled as $i=0,1,2, \ldots , N$, where $i=0$ is the oldest vertex. So the age of vertex $i$ is $a=N-i$.)  
Thus the probability that vertex $N+1$ attaches to a vertex of age $a$ is $P_N(a)=w(a)/W(N)$, where $W(N)=\sum_{1\leq a\leq N} w(a)$. To be more specific, we shall consider attachment kernels with an algebraic large age asymptotics,  
\begin{equation}
w(a) \sim A a^{-\mu} \quad{\rm when}\quad a\gg 1 
.
\label{e10}
\end{equation} 
For this $w(a)$, the normalization factor $W(N)$
converges when $\mu>1$ and diverges otherwise. Degree distributions in scale-free models of this kind were analysed in Ref.~\cite{dm00}. As exponent $\mu$ decreases from $+\infty$ to $-\infty$, the resulting network transforms from a chain to an ultimately compact star graph. Here we outline how the network varies with exponent $\mu$.  Details of derivations will be presented elsewhere \cite{DKM}.


\section{Results}
The simplest geometric characteristic of the network is the mean distance $\ell(N)$ of the $N$th vertex from the root. This quantity behaves similarly to the mean intervertex distance and to the diameter of the graph. We find that the mean depth of the last vertex scales as
\begin{equation}
\ell(N)
\cong 
\begin{cases}
C_1(\mu)\,\ln N                     & \mu <1 \cr
3\cdot \pi^{-2}\,(\ln N)^2       & \mu = 1 \cr
A^{-1}C_2(\mu)\,N^{\mu-1}           & 1<\mu < 2\cr
N/(A\ln N)                        & \mu=2\cr
N/W_1                             & \mu>2,
\end{cases} 
\label{e20}
\end{equation}   
where $W_1 = \sum_{a\geq 1}a w(a)$ is the first moment of the attachment rate
[$W_1$ is finite when $\mu>2$]. The amplitudes $C_1(\mu)$ and $C_2(\mu)$ in Eq.~\eqref{e20}
are 
\begin{equation}
C_1(\mu)=\frac{1}{\gamma{+}\psi(2{-}\mu)}, \ \  
\!C_2(\mu)=-\left[\frac{1}{\mu{-}1}+\frac{\pi}{\sin(\pi \mu)}\right]^{-1}\!\!\!,
\label{e30}
\end{equation}
where 
$\gamma\cong 0.5772\ldots$  is Euler's constant and $\psi(x)$ is the digamma function, $\psi(x)=\Gamma'(x)/\Gamma(x)$. So, for $\mu \leq 2$, the asymptotics of $\ell(N)$ is determined solely by the asymptotics of $w(a)$. Remarkably, if $\mu<1$, the result depends only on $\mu$. The mean depth of vertices in the network, $\langle \ell \rangle$, varies with $N$ similarly to $\ell(N)$, Eq.~\eqref{e20}, only corresponding coefficients differ when $\mu>1$.

Equation (\ref{e20}) shows that there are three different regimes: 
(i) when $\mu \leq 1$, the network is 
infinite-dimensional; 
(ii) when $1<\mu<2$, the network is $d$-dimensional, $d=1/(\mu{-}1)$ ($d$ is the Hausdorff dimension); 
(iii) when $\mu\geq 2$, the network is a one-dimensional object. The drastic changes take place at $\mu=1$ and $\mu=2$, when the zeroth moment 
$W_0\equiv W = \sum_{a\geq 1} w(a)$ and $W_1$ become divergent, respectively. 

The mean depth $\ell(N)$ is a basic measure of the size of the network. 
A more detailed characteristic is the depth distribution $Q_\ell(N)$, that is the mean number of vertices at distance $\ell$ from the root. We find a set of qualitatively different behaviors, and an additional special value $\mu=3$ where the second moment $W_2 = \sum_{a\geq 1}a^2 w(a)$ begins to diverge, demarcates different regimes. 
The major transitions \cite{note} in the behavior of the depth distribution still occur at $\mu=1$ and $\mu=2$: 
\begin{itemize}

\item[(i)]
when $\mu \geq 2$, this distribution is essentially a step function with a smeared front evolving as $\ell(N)$; 
 
\item[(ii)]
when $\mu \leq 1$, the distribution has a well-defined peak at $\sim \ell(N)$. 
\end{itemize}
The width of the front or peak scales as   
\begin{equation}
{\rm width} ~\propto~ 
\begin{cases}
(\ln N)^{1/2}                   & \mu<1 \cr
(\ln N)^{3/2}  &  \mu = 1 \cr
N^{\mu-1}          & 1<\mu<2\cr
N(\ln N)^{-3/2}         &\mu=2\cr
N^{(4-\mu)/2}               &2<\mu<3\cr
N^{1/2}(\ln N)^{1/2}             &\mu = 3\cr
N^{1/2}             &\mu >3 \,.
\end{cases} 
\label{e40}
\end{equation}

We also probed the degree distribution $n_k$. The dependence of this quantity on the exponent $\mu$ is less pronounced, namely the degree distribution exhibits an exponential decay when $\mu<1$ and a factorial decay ($\propto 1/k!$) when $\mu\geq 1$. The degree distribution has a surprising behavior at the point $\mu=1$ where the transition from 
an infinite- to a finite-dimensional geometry 
occurs. At this point 
\begin{equation}
n_k = \frac{1}{(k-1)!}e^{-1}
.
\label{e50}
\end{equation}
Intriguingly, the same degree distribution characterizes an {\em equilibrium} statistical ensemble of random trees \cite{b03}. 




\section{Mean depth}
In recursive trees, the addition of new vertices does not change the distances from the root of already existing vertices. This feature makes the problem analytically tractable. Consider first the general case when the attachment probability is 
proportional to 
an arbitrary function $w(i,k)$ of the birth time $k$ of a new vertex and of the birth time $i$ of an already existing vertex. Then the probability $\pi_\ell(j)$ that vertex $j$ is at distance $\ell$ from the root satisfies 
\begin{equation}
\pi_\ell(j) = 
\frac{1}{\sum_{i=0}^{j-1}w(i,j)}\sum_{i=0}^{j-1}w(i,j) \pi_{\ell-1}(i) 
, 
\label{e60}
\end{equation} 
$\pi_0(j)=\delta_{0,j}$, where the root vertex is $j=0$. 
Consequently for the average values of the integer powers $m$ of depths $\ell_i$ of individual vertices, we have 
\begin{equation}
\langle (\ell_j - 1)^m \rangle = 
\frac{1}{\sum_{i=0}^{j-1}w(i,j)}\sum_{i=0}^{j-1}w(i,j) \langle \ell_i^m \rangle
.
\label{e70}
\end{equation} 

Below we will discuss only $m=1$ and 
use the notation $\ell(i)\equiv \langle \ell_i \rangle$ for the mean depth of vertex $i$. 
In our networks $w(i,j)=w(j-i)$ and Eq.~\eqref{e70} gives 
\begin{equation}
\ell(N+1) = 1+\frac{1}{W(N)}\sum_{a=1}^N w(a)\, \ell(N+1-a)
.
\label{e90}
\end{equation} 
We do not solve this equation directly. Instead, we first guess a form of the solution and then verify its consistency. We start with a long-memory regime $\mu<1$. We know that for classical random recursive trees ($\mu=0$), the mean depth grows logarithmically with network size. Suppose that the same is valid when $\mu<1$, that is
\begin{equation}
\label{conj}
\ell(N)=C_1\ln N + D_1 +\ldots ,
\end{equation}
where $C_1$ and $D_1$ are some functions of $\mu$. To verify \eqref{conj} we first re-write recurrence \eqref{e90} as
\begin{equation}
\label{ell-recurr}
1=\frac{1}{W(N)}\sum_{a=1}^N w(a)\, [\ell(N+1)-\ell(N+1-a)]
.
\end{equation}
We then 
plug 
\eqref{conj} into \eqref{ell-recurr}, make the substitutions 
$w(a)\to Aa^{-\mu}$ and $W(N)\to A N^{1-\mu}/(1-\mu)$, 
and replace the summation by integration 
$\sum_{a=1}^N\to  N \int_0^1 dx$, $x=a/N$. 
After these transformations, Eq.~\eqref{ell-recurr} reduces to an equation that determines the amplitude
\begin{equation}
\label{C1-int}
C_1(\mu) = \left[-(1-\mu)\int_0^1 dx\,x^{-\mu}\ln(1-x)\right]^{-1}
.
\end{equation}
Computing the integral \eqref{C1-int} yields the resulting expression \eqref{e30} for $C_1(\mu)$. The correction $D_1(\mu)$ in Eq.~\eqref{conj} depends on the initial condition and therefore it cannot be computed in the realm of the continuous approach. 
Note the asymptotics: $C_1(\mu \to 1-0)\cong (6/\pi^2)\,(1-\mu)^{-1}$ and $C_1(\mu \to -\infty)\cong(\ln |\mu|)^{-1}$. 

In the marginal case $\mu=1$, 
$W(N) \cong A\ln N$. 
The divergence of the amplitude $C_1(\mu)$ as $\mu \to  1-0$ 
implies that in 
this 
case $\ell(N)$ grows faster than $\ln N$.  An algebraic growth is inconsistent with 
Eq.~\eqref{ell-recurr} where $W(N)$ is set to $\ln N$, 
while trying an accelerated logarithmic growth 
$\ell(N)\cong B(\ln N)^\lambda$  
we found consistency when
\begin{equation}
\label{Lambda}
\lambda=2, \quad 
B = -\left(2\int_0^1 \frac{dx}{x}\,\ln(1-x)\right)^{-1} 
= \frac{3}{\pi^2}
.
\end{equation}


When $\mu>1$, recurrence \eqref{e90} simplifies to 
\begin{equation}
\label{ell-recurrence}
1 = \sum_{a=1}^N w(a)\, [\ell(N)-\ell(N-a)]
\end{equation}
in the large $N$ limit. Expanding $\ell(N-a)$ into a Taylor series and keeping only the first  term of the expansion we recast \eqref{ell-recurrence} into a differential equation
\begin{equation}
\label{ell_N-ODE}
1 = \frac{d\ell}{dN}\sum_{a=1}^N a\,w(a)
.
\end{equation}
If the sum on the right-hand side converges ($\mu>2$), we have $d\ell/dN=1/W_1$  which leads to the announced expression of Eq.~\eqref{e20} in the regime $\mu>2$. The sum $\sum_{1\leq a\leq N} a\,w(a)$ diverges as $A\ln N$ when $\mu=2$. Solving $d\ell/dN=(A\ln N)^{-1}$ yields $\ell(N)=N/(A\ln N)$.   

When $1<\mu<2$, the sum on the right-hand side of Eq.~\eqref{ell_N-ODE} diverges as $N^{2-\mu}$,  so the asymptotics of the solution must be $\ell(N)\propto N^{\mu-1}$. 
Inserting $\ell(n)=A^{-1}C_2(\mu)\, n^{\mu-1}$ into \eqref{ell-recurrence} and replacing the sum by an integral 
we obtain 
\begin{equation*}
1=C_2(\mu)\int_0^1 dx\,x^{-\mu}\left[1-(1-x)^{\mu-1}\right]
,
\end{equation*}
which yields the resulting expression for $C_2(\mu)$ in Eq.~\eqref{e30}. 
Note the asymptotics: 
$C_2(\mu\to 2-0) \cong 1/(2-\mu)$ and $C_2(\mu\to 1+0) \cong (\pi^2/6)(\mu-1)$. 
Compare the latter with $C_1(\mu \to  1-0)\cong (6/\pi^2)\,(1-\mu)^{-1}$ which we obtained above.


\section{Depth distribution}
In the graph with aging, Eq.~\eqref{e60} takes the form: 
\begin{equation}
\label{pi_main}
\pi_{\ell+1}(n+1)=\sum_{i=1}^n \frac{w(n+1-i)}{W(n)}\,\pi_\ell(i)
.
\end{equation}
The average number of vertices at the $\ell^{\rm th}$ layer is equal to
\begin{equation}
\label{Q_pi}
Q_\ell(N)=\sum_{n=\ell+1}^N \pi_\ell(n)
.
\end{equation}
Equation~\eqref{pi_main} may be rewritten as 
\begin{equation}
\label{pi_eq}
0=\sum_{j=1}^{n-1} \frac{w(j)}{W(n-1)}\,[\pi_\ell(n-j)-\pi_{\ell+1}(n)]
.
\end{equation}
In the ``short memory'' regime ($\mu>1$), $W(n)\to W(\infty)$.  Since $w(j)$ quickly decays as $j$ increases, the major contribution to the sum on the right-hand side of 
Eq.~\eqref{pi_eq} 
occurs when $j$ is small. This prompts us to use the expansion 
$\pi_\ell(n-j)=\pi_\ell(n)-j\partial \pi_\ell(n)/\partial n +(1/2)j^2\,
\partial^2 \pi_\ell(n)/\partial n^2+\cdots$
and allows us to transform \eqref{pi_eq} into  
\begin{equation}
\left(\frac{\partial}{\partial \ell} + W_1\, \frac{\partial}{\partial n}\right)\! \pi_\ell(n)=\frac{1}{2}\left(W_2\,\frac{\partial^2}{\partial n^2}-\frac{\partial^2}{\partial \ell^2}\right)\! \pi_\ell(n)
\label{e300}
\end{equation}
The above derivation is valid when the moments $W_1$ and $W_2=\sum_{a\geq 1}a^2 w(a)$ are finite (that is, when $\mu>3$) and $\ell$ is large. 

To understand the form of the solution, it is sufficient to ignore the right-hand 
side of this equation, which leads to $\pi_\ell(n) = f(\ell -n/W_1)$, where $\int dx\,f(x)=1$. With Eq.~\eqref{Q_pi}, this results in $Q_\ell(N)=W_1[1-\theta(\ell-N/W_1)]$, where $\theta(x)$ is the Heaviside step function. Taking into account the right-hand 
side of Eq.~\eqref{e30} shows that this propagating front is, actually, smearing. In the front region we change variables 
$(n,\ell)\longrightarrow (t=n,x=\ell-n/W_1)$, $\pi_\ell(n)=p(t,x)$ 
and arrive to a diffusion equation,  
\begin{equation}
\label{diff}
\frac{\partial p}{\partial t} = D\,\frac{\partial^2 p}{\partial x^2}\,,\quad
D=\frac{W_2-W_1^2}{2W_1^3}
.
\end{equation}
Solving this equation leads to 
\begin{equation}
\label{Q_error}
Q_\ell(N) = \frac{W_1}{2}\,\,{\rm erfc}\left(\frac{N-W_1\ell}{W_1\sqrt{4DN}}\right)
,
\end{equation}
where ${\rm erfc}(x)= (2/\sqrt{\pi})\int_x^\infty dt\,e^{-t^2}$ is the complementary error function. 
The resulting width of the front is proportional to $\sqrt{4DN}$. This expression gives the front width growing as $\sqrt{N}$ when $\mu>3$ and also allows us to obtain estimates listed in Eq.~\eqref{e40} in the range $1<\mu\leq3$. In the latter region we simply take into account the dependence of the diffusion coefficient $D$, Eq.~\eqref{diff}, on $\mu$.

In the long memory regime ($\mu<1$), vertices of all ages continue to evolve and the appropriate variables are 
$\tau=\ln n$, $\xi=\ell-C_1\ln n$, $\pi_\ell(n)=p(\tau,\xi)$, 
with $C_1=C_1(\mu)$ given by Eq.~\eqref{e30}. 
In these variables, 
we obtain a diffusion equation, 
\begin{equation}
\label{Diff}
\frac{\partial p}{\partial \tau} = \frac{D\,C_1}{2}\,\frac{\partial^2 p}{\partial \xi^2}
,
\end{equation}
where $D=C_1^2\,(1-\mu)\int_0^1 dx\,x^{-\mu}[\ln(1-x)]^2-1$. The solution of Eq.~\eqref{Diff} is a Gaussian
which, after substitution into Eq.~\eqref{Q_pi}, gives 
\begin{equation}
\label{QN_sol}
Q_\ell(N) = \frac{N}{\sqrt{2\pi D\langle \ell\rangle}}\,
\exp\!\left[-\frac{(\ell-\langle\ell\rangle)^2}{2D\langle\ell\rangle}\right]
,
\end{equation}
where $\langle\ell\rangle=C_1\ln N$. 
This is a peak of width $\sim \langle\ell\rangle^{1/2}$ and not a step, in contrast to the short memory regime. 

Equation~\eqref{QN_sol} remains essentially valid in the marginal case $\mu=1$. 
Here $\langle\ell\rangle=3\cdot \pi^{-2}(\ln N)^2$, and the divergence $D\propto (1-\mu)^{-1}$ implies that one must replace $D$ by $\ln N$. Since the latter is proportional to $\langle\ell\rangle^{1/2}$ we conclude that 
the width of the peak in the marginal case is 
enhanced --- it scales as $\langle\ell\rangle^{3/4}$.


\section{Degree distribution}
The degree distribution can be written as
$N_k=\sum_{a=1}^N c_k(a,N)=N n_k$, where $c_k(a,N)$ is the average number of vertices of degree $k$ and age $a$. 
In the short memory regime, the joint distribution is (asymptotically) independent on the network size $N$, that is, $c_k(a,N)\to c_k(a)$. This simplifies calculations. On the other hand,  $c_k(a)$ changes only when the age is small, and hence we cannot use a continuum approximation. The equation for $c_k(a)$, 
\begin{equation}
\label{ck-rec}
c_k(a+1)=c_k(a) + w(a)[c_{k-1}(a)-c_k(a)]
,
\end{equation}
can be solved and it leads \cite{DKM} to the following 
universal asymptotics of $n_k$: 
\begin{equation}
\label{nk-asymptotic}
n_{k+1}\propto \frac{1}{k!}\quad{\rm as}\quad k\to\infty
.
\end{equation}
This asymptotics is valid for the general class of attachment kernel~\eqref{e10}, where $\mu \geq 1$. 

In the evolving regime ($\mu<1$), the density $c_k(a,N)$ continues to evolve as the network size grows. In this case we employ a continuum approximation. The governing equation for $c_k(a,N)$ reads 
\begin{equation}
\label{ckN}
\left(\frac{\partial}{\partial a}+\frac{\partial}{\partial N}\right)c_k
=\frac{w(a)}{W(N)}\,(c_{k-1}-c_k)
.
\end{equation} 
When $\mu<1$, the analysis \cite{DKM} of Eq.~\eqref{ckN} gives an exponential asymptotics of the degree distribution: 
\begin{equation}
n_k=\frac{e^{-\gamma-\psi(1-\mu)}}{1-\mu}
\left(\frac{1-\mu}{2-\mu}\right)^k\quad {\rm when}  
\quad k\gg 1
.
\end{equation}
In the marginal case ($\mu=1$),  an inverse factorial form of the degree distribution, Eq.~\eqref{e50}, is valid for all $k$.  





\section{Discussion and summary}
Several network features are puzzling and deserve further attention. 

(i) 
The degree distribution of our growing random graph at $\mu=1$ is the same as for the static ensemble of random trees \cite{b03}. We have no explanation for this surprising coincidence. The two networks differ dramatically: our graph is a small world at $\mu=1$, while according to  Ref.~\cite{rs67} a tree which is randomly chosen from an ensemble of all trees is 
finite-dimensional 
with a high probability 
(the Hausdorff dimension of such trees is $d=2$). Interestingly, at 
the second transition point $\mu=2$, the degree distribution does not exhibit any special behavior. 

(ii) 
The asymptotic 
$\langle \ell \rangle_N \sim (\ln N)^2$ 
arises at a searchability point of specific networks discussed in Refs.~\cite{k00,k06} in the context of decentralized search algorithms. These
networks are $d$-dimensional lattices with random shortcuts which connect
vertex pairs with probability decaying as a power-law of an Euclidean separation. 
The greedy algorithms work rapidly at some value of the decay exponent --- the searchability point. At that point, the run time of these algorithms is of the order of $\langle \ell \rangle_N$. 
The coincidence of our result with that of Refs.~\cite{k00,k06} suggests that the tree approximation is valid at the searchability point of (some) networks with loops.  

We have studied simplest recursive graphs which have rapidly decreasing degree distributions. Yet similar phenomena should be observed in growing networks with more complex (e.g., skewed) degree distributions.  Equations~\eqref{e60} and \eqref{e70} allow one to analyse recursive trees with arbitrary kernels $w(i,j)$ generating various degree distributions.

In summary, we have demonstrated a rich set of regimes in a global organization of growing networks with aging. Our results have been derived in the context of a specific model, yet the major conclusions should be applicable to a larger class of networks. 
Our analysis has exploited the absence of loops. Accounting for loops is a challenging direction for future work. 

\acknowledgments
This work was partially supported by projects POCTI: FAT/46241, MAT/46176,
FIS/61665 and BIA-BCM/62662, and DYSONET. We thank A.~V. Goltsev, M. Hase, M.~E.~J. Newman, A.~N. Samukhin, and B.~Waclaw for useful discussions.



\begin{thebibliography}{0}


\bibitem{ws98}  
  \Name{Watts~D.~J. \and Strogatz~S.~H.}
  \REVIEW{Nature}{393}{1998}{409}.

\bibitem{k00}
  \Name{Kleinberg J.~M.}
  \REVIEW{Nature}{406}{2000}{845}.

\bibitem{k06}
  \Name{Kleinberg~J.} 
  \Book{Proceedings of the International Congress of Mathematicians (ICM), Madrid,
August 22--30, 2006}
  \Editor{M. Sanz-Sol\'e, J. Soria, J.~L. Varona \and J. Verdera}
  \Publ{EMS Publishing House, Z\"urich}
  \Year{2007}
  \Page{1019}.
  
\bibitem{m02}
  \Name{Moukarzel~C.~F. \and Argollo de Menezes M.}
  \REVIEW{Phys. Rev. E}{65}{2002}{056709}.

\bibitem{h07}
  \Name{Holme P.}
  \REVIEW{Physica A}{377}{2007}{315}. 

\bibitem{hb06}
\Name{Hinczewski~M. \and Berker~A.~N.} 
\REVIEW{Phys. Rev. E}{73}{2006}{066126}. 

\bibitem{tzs06}
\Name{Tian~L., Zhu~C.-P., Shi~D.-N., Gu~Z.-M. \and Zhou~T.}  
\REVIEW{Phys. Rev. E}{74}{2006}{046103}.  

\bibitem{zzw07}
\Name{Zhang~Z., Zhou~S., Wang~Z. \and Shen~Z.} 
\REVIEW{J. Phys. A}{40}{2007}{11863}; 
\Name{Zhang~Z., Zhou~S., Shen~Z. \and Guan~J.} 
\REVIEW{Physica A}{385}{2007}{765}. 

\bibitem{l07}
\Name{Lambiotte~R.}  
\REVIEW{J. Stat. Mech.}{$\!\!$}{2007}{P02020}. 

\bibitem{mh} 
  \Name{Smythe~R.~T. \and Mahmoud H.}
  \REVIEW{Theor. Probab.\ Math.\ Statist.}{51}{1995}{1}.

\bibitem{m74}
  \Name{Moon~J.~W.}
  \Book{London Mathematics Society Lecture Notes Series}
  \Vol{13}
  \Publ{Cambridge University Press, Cambridge}
  \Year{1974}
  \Page{125}. 

\bibitem{s90}
  \Name{Szyma\'nski~J.}
  \REVIEW{Theor. Comp. Sci.}{74}{1990}{355}.

\bibitem{m91}
  \Name{Mahmoud H.}
  \REVIEW{Probab. Eng. Info. Sci.}{5}{1991}{5}.

\bibitem{d96}
  \Name{Dobrow~R.~P.}
  \REVIEW{J. Appl. Prob.}{33}{1996}{749}.

\bibitem{drmota} 
  \Name{Drmota~M. \and Gittenberger~B.}
  \REVIEW{Random Struct.\ Alg.}{10}{1997}{421}; 
  \Name{Drmota~M. \and Hwang~H.-K.} 
  \REVIEW{Adv.\ Appl.\ Probab.}{37}{2005}{321}.


\bibitem{GNR} 
  \Name{Krapivsky~P.~L. \and Redner~S.}
  \REVIEW{Phys.\ Rev.\ E}{63}{2001}{066123}; 
  \REVIEW{Phys.\ Rev.\ Lett.}{89}{2002}{258703}.

\bibitem{dm00}
  \Name{Dorogovtsev~S.~N. \and Mendes J.~F.~F.}
  \REVIEW{Phys. Rev. E}{62}{2000}{1842}.
 
\bibitem{DKM}
  \Name{Dorogovtsev~S.~N., Krapivsky~P.~L. \and Mendes J.~F.~F.} in preparation.

\bibitem{note}
  These transitions are different from critical phenomena; for a discussion of 
  this subtlety in the context of networks, see 
   \Name{Dorogovtsev~S.~N., Goltsev~A.~V. \and Mendes J.~F.~F.}
  \REVIEW{arXiv:0705.0010 [cond-mat]}{\!\!}{2007}\!\!.
 
\bibitem{b03}
\Name{Bialas~P., Burda~Z., Jurkiewicz~J. \and Krzywicki~A.} 
\REVIEW{Phys. Rev. E}{67}{2003}{066106}.

\bibitem{rs67}
\Name{R\'{e}nyi~A. \and Szekeres G.} 
\REVIEW{J. Austral.\ Math.\ Soc.}{7}{1967}{497}; 
\Name{G. Szekeres} 
in: {\em Combinatorial Mathematics X, 1982}, 
{\em Lecture Notes in Mathematics}, Vol. {\bf 1036} 
(Springer-Verlag, Berlin), 1983, p.~392. 
  

 









\end{thebibliography}
\end{document}